\documentclass[11pt]{article}
\usepackage{amsmath}
\usepackage{axodraw}

\newcommand{\Ref}[1]{(\ref{#1})}
\newcommand{\chiM}[4]{
\begin{array}{r}{\scriptstyle #2}\\{\scriptstyle #1}\end{array}\!\! \chi\!\! 
\begin{array}{l}{\scriptstyle #4}\\{\scriptstyle #3}\end{array}}

\newcommand{\phiM}[4]{
\begin{array}{r}{\scriptstyle #2}\\{\scriptstyle #1}\end{array}\!\! \phi\!\! 
\begin{array}{l}{\scriptstyle #4}\\{\scriptstyle #3}\end{array}}

\begin{document}

\title{Contraction and decomposition matrices for vacuum diagrams}
\author{K. Knecht\thanks{%
Email: Kenny\_Knecht@applied-maths.com}, H. Verschelde\thanks{%
Email: Henri.Verschelde@rug.ac.be} \\
Department of Mathematical Physics and Astronomy,\\ University of Ghent,\\
Krijgslaan 281, 9000 Ghent, Belgium}
\maketitle

\begin{abstract}
Tensor reduction of vacuum diagrams uses contraction and decomposition matrices.
We present general recurrence relations for the calculation of those matrices and an explicit 
formula for the 3-loop decomposition matrix and its determinant.
\end{abstract}

\newpage 
\section{Introduction}
This letter is mainly conceived as an extension of the two-loop results 
for contraction and decomposition matrices of \cite{Davy4}. 
For a list of many usefull applications and other literature on this matter we gladly refer 
the reader to this paper and the references therein. Here we would only like to point
out that the higher loop results are also very important.
 For instance they can be applied in the calculations
of the anomalous dimensions (currently 3 and 4 loops are feasable \cite{chet,QCD}) 
or the moments of deep inelastic 
structure functions (cfr. 3-loop calculation in \cite{Larin:QCD3inelast}).

The remainder of the article is organized is follows. In section \ref{notat} 
we introduce the notation. Although
we tried to follow the notations of \cite{Davy4} as closely as possible, 
we had to make some adaptions to
be able to describe the tensor structure of a general $L$-loop diagram. In section \ref{recur} the general
recurrence relations are presented and in section \ref{explic} an explicit solution in the 3-loop case is
derived. The last section is a summary and conclusion.

\section{Notations}\label{notat}
In general a $L$-loops diagram has $N_i$ tensorindices in every loop $i= 1, \ldots,L$.
 This gives us the index-set
$I = \bigcup_{i} I_i$, $I_i = \{\mu_{i,1},\mu_{i,2},\ldots,\mu_{i,N_i}\}$ for every loop $i$. A vacuum diagram 
$G[I]$ with this tensorstructure can be
decomposed in all the possible products of metric tensors we can produce  from the index-set $I$, i.e. in a shorthand
we get
\begin{equation}
G[I] =  \sum_{\sigma(I)} a_{\sigma}\prod_{j=1}^{N/2} g_{\sigma(j) \sigma(j+1)} \label{shorthand}
\end{equation}
We always assume the $N = \sum_i N_i$ is even otherwise the  vacuumdiagram $G[I]$ is identically zero.
Many of these coefficients $a$ will have the same value: the left-hand side of the equation is symmetrical
in the indices of $I_i$, thus so should the right-hand side. Now if we consider the metric tensor
to be an object which makes a connection, either between two different loops
 ($g_{\mu_{i,j}\mu_{j,k}}$) or within 
the same loop ($g_{\mu_{i,j}\mu_{i,k}}$), it is
easy to see that the only distinct values $a_\sigma$ will correspond to a different number of connections
between the loops. Therefore we introduce an object which we call a \emph{link}  and contains this information. It 
is characterized by the number of connections $t_{ij}$ between loop $i$ en $j$ ($L(L+1)/2$ numbers). It might also
be characterized by the number of metric tensors $s_i$ which stay within a certain loop ($L$ numbers). We have the 
following
relations
\begin{equation}
2 s_i+\sum_j t_{ij} = N_i, \forall i \in {1,2,\ldots,L}.
\end{equation}
The tensorconfiguration $S_l$ is the symmetric sum of products of metric tensors which belong to a certain link
$l$. The number of terms in such a tensorconfiguration is
\begin{equation}
c_l=\frac{\prod N_i!}{\prod (2 s_i)!! t_{ij}!}.\label{eq:ctt:cl}
\end{equation}
Now we can write \Ref{shorthand} as 
\begin{equation}
G[I] = \sum_l a_l S_l\label{ontbind}
\end{equation}
where sum runs over all possible links for the tensorconfiguration $I$.
We get a system of equations by contracting this expression with each $S_{l'}$: 
\begin{equation}
G[I]\otimes S_{l'}= G[I]^{(l')} = \sum_l a_l S_l\otimes S_{l'}=\sum_l a_l  \chi_{ll'}
\end{equation}
which is the definition of the contraction matrix $\chi$. Note that we write an $\otimes$ if tensors are involved. 
This is can be inverted to
\begin{equation}
a_l = (\chi^{-1})_{ll'}G[I]^{(l')} = \phi_{ll'}G[I]^{(l')}.
\end{equation}
where $\phi$ is the decomposition matrix. In this definition both matrices are symmetric. We use the notation
$\chi_{ll'}$ and $\phi_{ll'}$ if  the link is completely general. Otherwise we shall write
\begin{equation}
\chiM{(t)}{(s)}{(t')}{(s')}.
\end{equation}
Here $s$ is the columnmatrix containing the number $s_i$ of metric
tensors which stay within the loop $i$ and $t$ is the triangular matrix
containing the number $t_{ij}$ of metric tensors which connect loops $i$ and $j$.
Here we will not write the corresponding indices though. So if we write 
$(s)$, we really mean $(s_k)$ and $(s-\delta_i)$ is really $(s_k-\delta_{ik})$, 
$k$ running from 1 to $L$ and analogous
for $(t)$.

In order to get used to this notations we will derive a simple identity which we will use further on.
The explicit solution of this problem for the one-loop case was allready known in \cite{veltman}:
\begin{equation}
^{(N/2)}\chi^{(N/2)}= (N-1)!! 2^{N/2} (d/2)_{N/2} \label{eq:ctt:1luschi}
\end{equation}
Now if we have a general $L$-loop diagram and sum over the tensorconfigurations
$
\sum_l S_l 
$ this  equals the tensorconfiguration of the 1-loop diagram with $I^{\rm 1-loop} = \cup_i I_i$.
If $r_l$ is an arbitrary term form $S_l$, we have using the symmetry of $\chi$
\begin{equation}
\begin{split}\sum_{l'}\chi_{ll'} &=  S_l \otimes \sum_{l'}S_{l'} = c_l r_l \otimes \sum_{l'}S_{l'} \\&=
\frac{c_l}{(N-1)!!}\sum_{l}S_{l} \otimes \sum_{l'}S_{l'}  =\frac{c_l}{(N-1)!!}S_{\rm 1-loop} \otimes S_{\rm 1-loop} 
\\&= c_l 2^{N/2} (d/2)_{N/2}
\end{split}\end{equation}
using \Ref{eq:ctt:1luschi}. If we multiply by $\phi$, we obtain
\begin{equation}
\sum_{l'}\phi_{ll'}c_{l'} =  \frac{1}{ 2^{N/2} (d/2)_{N/2}}\label{ident}
\end{equation}
Here we have derived this expression on very general grounds, while the authors of \cite{Davy4} have
proved it by explicit 2-loop calculations. Like these authors we will use identity \Ref{ident}
in order to obtain explicit solutions for the recurrence relations we will now derive.

\section{Recurrence relations}\label{recur}
Now we can construct recurrence relations for a  contraction tensor. We can do 
this using a similar method as in 
\cite{Davy4} by writing $S_l$ as a partial derivative of tensors. 
However we found it to be very convient to add 
the metric tensors one by one in $S_l\otimes S_{l'}=  \chi_{ll'}$ in
 a grafic represenation \cite{thesis}. Either way we get the 
following
recurrence relations for an $L$-loop contraction tensor
\begin{eqnarray} 
\lefteqn{\chiM{(t)}{(s)}{(t')}{(s')} =\frac{N_i N_j}{t'_{ij}}\biggl[
(d+N_i+N_j-t_{ij}-1)\chiM{(t-\delta_{ij})}{(s)}{(t'-\delta_{ij})}{(s')}}\nonumber\\&&
+t_{ij}\chiM{(t+\delta_{ij})}{(s-\delta_{i}-\delta_{j})}{(t'-\delta_{ij})}{(s')}
+\sum_{k\neq 
i,j}t_{ik}\chiM{(t-\delta_{jk}+\delta_{ik})}{(s-\delta_{i})}{(t'-\delta_{ij})}{(s')}
\nonumber\\&&
+\sum_{k\neq i,j}t_{jk}
\chiM{(t-\delta_{ik}+\delta_{jk})}{(s-\delta_{j})}{(t'-\delta_{ij})}{(s')}
+\sum_{k\neq i}\sum_{l\neq j,k}t_{kl}
\chiM{(t-\delta_{ik}-\delta_{jk}+\delta_{kl})}{(s)}{(t'-\delta_{ij})}{(s')}\nonumber\\&&
 +\sum_{k\neq i,j}2s_{k} 
\chiM{(t-\delta_{ik}-\delta_{jk})}{(s+\delta_{k})}{(t'-\delta_{ij})}{(s')}\biggr],\label{chirecur1}
\end{eqnarray}
and
\begin{eqnarray} 
\lefteqn{\chiM{(t)}{(s)}{(t')}{(s')} =
\frac{N_i(N_i-1)}{2s'_{i}}\biggl[(d+2N_i - 2s_i-2) 
\chiM{(t)}{(s-\delta_{i})}{(t')}{(s'-\delta_{i})}}\nonumber\\&&
+\sum_{k\neq 
i}\sum_{l<k}2t_{kl}\chiM{(t-\delta_{ik}-\delta_{jk}+\delta_{kl})}{(s)}{(t')}{(s
'-\delta_{i})} +\sum_{k\neq i}2s_{k}
\chiM{(t-2\delta_{ik})}{(s+\delta_{k})}{(t')}{(s'-\delta_{i})}\biggr].\label{chirecur2}
\end{eqnarray}
These expressions for $\chi$ induce similar relations for $\phi$. We get
\begin{eqnarray} 
\lefteqn{t'_{ij} \phiM{(t-\delta_{ij})}{(s)}{(t'-\delta_{ij})}{(s')} = N_i 
N_j\biggl[(d-1+N_i+N_j-t_{ij})\phiM{(t)}{(s)}{(t')}{(s')}}\nonumber\\&&
+(t_{ij}-1)\phiM{(t-2\delta_{ij})}{(s+\delta_{i}+\delta_{j})}{(t')}{(s')}
+\sum_{k\neq 
i,j}t_{ik}\phiM{(t+\delta_{jk}-\delta_{ik}-\delta_{ij})}{(s+\delta_{i})}{(t')}{(s')}\nonumber\\&&
+\sum_{k\neq 
i,j}t_{jk}\phiM{(t+\delta_{ik}-\delta_{jk}-\delta_{ij})}{(s+\delta_{j})}{(t')}{(s')}
\nonumber\\&
&
+\sum_{k\neq i}\sum_{l\neq 
j,k}t_{kl}\phiM{(s)}{(t+\delta_{ik}+\delta_{jk}-\delta_{kl}-\delta_{ij})}{(t')}{(s')}
\nonumber\\&&
 +\sum_{k\neq i,j}2s_k\phiM{(t-\delta_{ik}-\delta_{jk})}{(s+\delta_{k})}{(t')}{(s')}\biggr]
\label{eq:contr:phi3t}
\end{eqnarray}
and
\begin{eqnarray} 
\lefteqn{ (2 s_i) \phiM{(t)}{(s-\delta_{i})}{(t')}{(s'-\delta_{i})}=  N_i(N_i-1)[(d+2N_i-2 
s_i-2) 
\phiM{(t)}{(s)}{(t')}{(s')}}\nonumber\\&&
+\sum_{k\neq i}\sum_{l<k}2 
t_{ik}\phiM{(t+\delta_{ik}+\delta_{jk}-\delta_{kl})}{(s-\delta_{i})}{(t')}{(s')}
 +\sum_{k\neq i}2s_k 
\phiM{(t+2\delta_{ik})}{(s-\delta_{k}-\delta_{i})}{(t')}{(s')}].\label{eq:contr:phi3s}
\end{eqnarray}
The expressions \Ref{chirecur1} and \Ref{chirecur2} are very usefull when contsructing the matrices
through recurrence relations, while \Ref{eq:contr:phi3s} and \Ref{eq:contr:phi3t} are not
fit for direct use: they will however allow us to construct explicit solutions for the 3-loop case.
We have tested these recurrence relations by comparing their results with a {\tt Mathematica}-package
by Misiak \cite{mis} which uses the actual $S_l \otimes S_{l'}$ contraction.

\section{Explicit solutions}\label{explic}
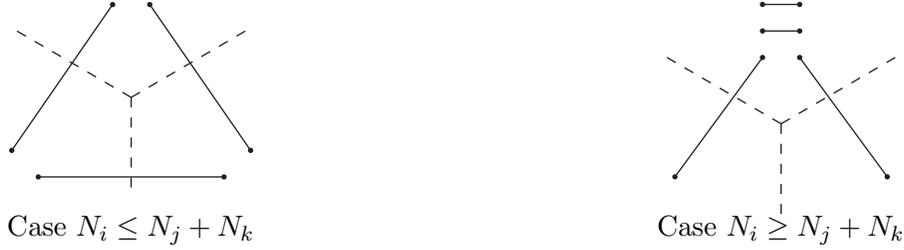
\begin{figure}
\begin{center}
\begin{picture}(100,100)(0,-10)
\DashLine(50,50)(50,16){4}
\DashLine(50,50)(93,75){4}
\DashLine(50,50)(7,75){4}
\Vertex(43,85){1}
\Vertex(5,30){1}
\Vertex(57,85){1}
\Vertex(95,30){1}
\Vertex(85,20){1}
\Vertex(15,20){1}
\Line(43,85)(5,30)
\Line(57,85)(95,30)
\Line(85,20)(15,20)
\Text(50,0)[]{Case $N_i\leq N_j +N_k$}
\end{picture}\hspace{50mm}
\begin{picture}(100,100)
\DashLine(50,50)(50,16){4}
\DashLine(50,50)(93,75){4}
\DashLine(50,50)(7,75){4}
\Vertex(43,95){1}
\Vertex(57,95){1}
\Vertex(43,85){1}
\Vertex(57,85){1}
\Vertex(43,75){1}
\Vertex(57,75){1}
\Vertex(10,30){1}
\Vertex(90,30){1}
\Line(43,95)(57,95)
\Line(43,85)(57,85)
\Line(43,75)(10,30)
\Line(57,75)(90,30)
\Text(50,10)[]{Case $N_i\geq N_j +N_k$}
\end{picture}
\end{center}
\caption{The different 3-loop cases}
\label{fg:ctt:3lusgevallen}
\end{figure}

Like in \cite{Davy4} we will try to reduce the decomposition matrix to a unique value, which we will
then compute by using relation \Ref{ident}. In the 3-loop cases however, we have  two classes, 
each of them with 
different endpoints for recurrence relations. 
They are shown in figure \ref{fg:ctt:3lusgevallen} (lines that connect 
vertices are metric tensors, dashed lines separate the loops)
\\

We start from the simple case where $N_i\geq N_j +N_k$ and
$s'_j= 
s'_k= t'_{jk}=0$.
By applying \Ref{eq:contr:phi3s} 
 on $s_i$ we obtain the following endpoint of the recurrence relations: 
$t'_{ij}=t_{ij}=N_j$  and $t'_{ik}=t_{ik}=N_k$:
\begin{eqnarray}
\lefteqn{^{((N_i-N_j-N_k)/2,0,0)}_{(0,N_j,N_k)}\phi^{(s)}_{(t)}}\nonumber\\&=&
\frac{(-)^{s_i+(N_i-N_j-N_k)/2}s_i!(N_j+N_k)!(d/2-1+N_j+N_k)}
{N_i! (d/2-1+N/2-s_i)_{s_i+1}} \nonumber\\
&& 
\times\,^{(0,0,0)}_{(0,N_j,N_k)}\phi^{(0,0,0)}_{(0,N_j,N_k)}\label{eq:contr:firstres3l}
 \end{eqnarray}
Calculating the normalisation factor $ 
^{(0,0,0)}_{(0,N_j,N_k)}\phi^{(0,0,0)}_{(0,N_j,N_k)}$ 
with the aid of \Ref{ident} gives 
\begin{equation}
\sum_{s_i}\sum_{s_j}\sum_{s_k}\, 
^{((N_i-N_j-N_k)/2,0,0)}_{(N_j,N_k,0)}\phi^{(s)}_{(t)}
\frac{N_i!N_j!N_k!}{2^{\Sigma s}s_i!s_j!s_k! 
t_{ij}!t_{ik}!t_{jk}!}=\frac{1}{2^{N}(d/2)_N}
\label{eq:contr:phisum3l}
\end{equation}
Because \Ref{eq:contr:firstres3l} is only dependent on $s_i$ we shall evaluate the other 2 summations first. If we 
add a factor $(N_i-2s_i)!$ in the numerator  we obtain
\[
\sum_{s_j}\sum_{s_k}\frac{(N_i-2s_i)!N_j!N_k!}{2^{s_j+s_k}s_j!s_k! 
t_{ij}!t_{ik}!t_{jk}!}.
\]
We recognize the expression \Ref{eq:ctt:cl} for counting the number of terms in a
tensorconfiguration $S_l$ with the following number of tensorindices $(N_i-2s_i,N_j,N_k)$. By summing over 
all possible $s_j$ and $s_k$ we can readily see that we really obtain the number of terms in 
a 2-loop diagram with  $(N_i-2s_i,N_j+N_k)$ tensorindices. This gives us
\begin{equation}
\sum_{s_j}\sum_{s_k}\frac{(N_i-2s_i)!N_j!N_k!}{2^{s_j+s_k}s_j!s_k! 
t_{ij}!t_{ik}!t_{jk}!}=
\frac{(N_j+N_k)!}{2^{(N_j+N_k-N_i+2s_i)/2}(\frac{N_j+N_k-N_i+2s_i}{2})!}
\end{equation}
The remaining summation over $s_i$  is identical to the expression we get by inserting (25) in (26) of 
\cite{Davy4} which can be summed up to a hypergeometric function. Eventually we get 
\begin{equation}
^{(0,0,0)}_{(N_j,N_k,0)}\phi^{(0,0,0)}_{(N_j,N_k,0)}=\frac{(\frac{d-2}{2})_{
N_j+N_k}}{(N_j+N_k)!
(d-2)_{N_j+N_k}
(\frac{d}{2})_{N_j+N_k}}\label{eq:contr:result1}
\end{equation}
Note that we have not explicitely used the fact that there are only 3 loops: this result is valid for
every diagram with $N_1\geq \sum_{i=2}^L N_i$.

If $N_i\leq N_j +N_k,\forall i$ again we will start from the simpler 
case $^{(0)}_{(t_0)}\phi^{(s)}_{(t)}$, where $t_0$ stands for
 $t_{ij}= \frac{N_i+N_j-N_k}{2}, \forall i\neq j$. 
Applying the recurrence relation \Ref{eq:contr:phi3s} in this case  
 eventually we will end up in a situation where
$^{(0)}_{(t_0)}\phi^{(s)}_{(t)}$ is completely expressed as a function of one 
 unique unknown factor $^{(0)}_{(t_0)}\phi^{(0)}_{(t_0)}$, which thus fulfils the role
of a  normalisation factor. The result of the recurrence relation \Ref{eq:contr:phi3s} is
\begin{eqnarray} 
\lefteqn{^{(0)}_{(t_0)}\phi^{(s)}_{(t)}=\sideset{}{_{m=\max(0,s_i-s_j)}^{s_i}}\sum
\sideset{}{_{n=\max(0,m-s_k)}^{\min(m,t_{jk})}}\sum
\sideset{}{_{p=\max(0,s_j-s_i-s_k+2m-n)}^{\min(s_j-s_i+m,t_{ik}+2m-n)}}\sum}\nonumber\\
&&  \frac{s_i!}{(s_i-m)!(m-n)!n!}\frac{s_j!}{(s_j-s_i+m-p)!p!}
\frac{s_k!}{(s_k-sm+n-s_j+s_i+p)!}\nonumber
\\&& (-1)^{s_i+s_k-m+n+p}\frac{t_{jk}!}{(t_{jk}-n)!}
\frac{(t_{ik}+2m-n)!}{(t_{ik}+2m-n-p)!}\frac{(t_{ij}+2s_i-2m+n+p)!}{(t_{ij}+
s_i+s_j-s_k)!}\nonumber\\&&
(d/2-1+N_i-s_i)_{s_i}(d/2-1+N_j-s_j+s_i-m)_{s_j-s_i+m}\nonumber
\\&&(d/2-1+N_k-(s_k+s_j+s_i-2m+n+p))_{s_k+s_j+s_i-2m+n+p}
\nonumber\\&&
\,^{(0)}_{(t_0)}\phi^{(0)}_{(t_0)}.\label{eq:contr:tssres}
\end{eqnarray}
In order to calculate the normalisation factor we have to sum up this 
expression \`a la \Ref{eq:contr:phisum3l}, 
which gives us a nine-fold summation! It is clear that
 this is practically a dead-end street. Nevertheless by making connection 
with the
simpler case $N_i\geq N_j +N_k$ we will be able to prove
 the following lemma:
\begin{equation}
\phiM{(t_0)}{(0)}{(t_0)}{(0)} \stackrel{\rm ?}{=}f(N_i,N_j,N_k)= 
\frac{(d-2)_{N/2}(\frac{d}{2})_{N/2}\prod_i N_i! 
(d/2+N_i-1)_{N/2-N_i}}
{\prod_i (N/2-N_i)!(d/2+N/2-N_i-1)_{N_i}}.\label{eq:contr:toprove}
\end{equation}
We will prove by induction. In the  borderline case $N_i=N_j+N_k$  \Ref{eq:contr:toprove} 
reduces to
\Ref{eq:contr:result1}, which has been proven. This is the starting point of our induction. In order to proceed we 
must find
a relation between $\phiM{(t-\delta_{ij})}{(0)}{(t-\delta_{ij})}{(0)}$ and 
$\phiM{(t)}{(0)}{(t)}{(0)}$ (we now no longer write the index 0). In order to establish this connection
we use \Ref{eq:contr:phi3t} 
\begin{eqnarray*} 
\lefteqn{t_{ij} \phiM{(t-\delta_{ij})}{(0)}{(t-\delta_{ij})}{(0)}}\\& =& N_i 
N_j[(d+N_i+N_j-t_{ij}-1)\phiM{(t)}{(0)}{(t)}{(0)} 
+(t_{ij}-1)\phiM{(t-2\delta_{ij})}{(\delta_{i}+\delta_{j})}{(t)}{(0)}\\
&&
+t_{ik}\phiM{(t+\delta_{jk}-\delta_{ik}-\delta_{ij})}{(\delta_{i})}{(t)}{(0)}
+t_{jk}\phiM{(t+\delta_{ik}-\delta_{jk}-\delta_{ij})}{(\delta_{j})}{(t)}{(0)}.
\end{eqnarray*}
and sequentially \Ref{eq:contr:phi3s} 
\begin{eqnarray} 
\lefteqn{ \phiM{(t-\delta_{ij})}{(0)}{(t-\delta_{ij})}{(0)} =\frac{N_i N_j}{t_{ij}}\left[ 
(d+N_i+N_j-t_{ij}-1) 
-\frac{(t_{ij}-1)}{d/2-2+N_i}\right.}\nonumber\\&&+\frac{(t_{ij}-1)t_{jk}(t_{ik}+1)}{(
d/2-2+N_i)(d/2-2+N_j)}-\frac{t_{ik}(t_{jk}+1)}{d/2-2+N_i}\nonumber\\&&
\left.-\frac{t_{jk}(t_{ik}+1)}{d/2-2+N_j}\right]\,_{(t)}^{(0)}\phi_{(t)}^{(0)},\label{recurs1N}
\end{eqnarray}
$f(N_i,N_j,N_k)$ on the other hand satisfies the following recurrence relation 
\begin{eqnarray} 
\lefteqn{f(N_i-1,N_j-1,N_k)= \frac{N_i N_j (d/2-2+N/2-N_k)}{(\frac{N_i 
+N_j-N_k}{2})
(d/2-2+N_i)(d/2-2+N_j)}}\nonumber\\&&\times(d/2-2+N/2)(d/2-3+N/2)f(N_i,N_j,N_k)
\label{recurs2N}
\end{eqnarray}
By substituting $t_{ij}= \frac{N_i+N_j-N_k}{2}, \forall i \neq j$ it is 
easy to establish the equality
of \Ref{recurs1N} and \Ref{recurs2N} so not only  the lemma
 $^{(0)}_{(t_0)}\phi^{(0)}_{(t_0)}\equiv f(N_i,N_j,N_k)$ is proven but also 
the hideous expressions \Ref{eq:contr:tssres}.

If we start from the general case $_{(t')}^{(s')}\! 
\phi_{(t)}^{(s)}$ we can first apply
 \Ref{eq:contr:phi3s} in order to reduce $s'_1,s'_2$ and $s'_3$ to zero. 
This recurrence relation 
can be solved explicitely (see \cite{Davy4}). Together with 
\Ref{eq:contr:tssres} and \Ref{eq:contr:toprove} we have
established an explicit form for 
$_{(t')}^{(s')}\! 
\phi_{(t)}^{(s)}$. After careful substitutions we can write this in a symmetrical form
\begin{eqnarray}
\lefteqn{_{(t')}^{(s')}\! \phi_{(t)}^{(s)}= \sum_{l''} \sum_{l'''} 
\sum_{l^{v}} 
\frac{\prod [(-)^{s''_i+s_i^v}s_i!s'_i!]}{\prod[ s'''_i! N_i! 
(s_i^v-s'''_i)!(s''_i-s'''_i)!]} 
\frac{\prod 
t'''_{ij}!}{\prod[(\frac{t''_{ij}-t_{ij}}{2})! 
(\frac{t^v_{ij}-t'_{ij}}{2})!]}}
\nonumber\\&&
\times 
\frac{t_{jk}!}{(t_{jk}-s''_i+s_i''')!}\frac{t'_{jk}!}{(t'_{jk}-s^v_i+s_i''')
!}
\frac{(t_{ik}+s''_i-s_i''')!}{(t_{ik}+s''_i-s_i'''-s''_j+s_j''')!}
\nonumber\\&&\times
\frac{(t_{ik}+s^v_i-s_i''')!}{(t_{ik}+s^v_i-s_i'''-s^v_j+s_j''')!}
\frac{(t_{ij}+s''_i-s_i'''+s''_j-s_j''')!}{t'''_{ij}!}
\nonumber\\&&\times
\frac{(t_{ij}+s^v_i-s_i'''+s^v_j-s_j''')!}{t'''_{ij}!}
 \frac{\prod (\frac{d-2}{2}+t'''_{ij})_{t'''_{ik}+t'''_{jk}}}{(d-2)_{\Sigma 
t'''_{ij}}(d/2)_{\Sigma t'''_{ij}}}
\nonumber\\&&\times 
\frac{(\frac{d-2}{2})_{(t_{ij}+t_{ik}+t'''_{ij}+t'''_{ik})/2}
(\frac{d-2}{2})_{(t''_{ij}+t''_{jk}+t'''_{ij}+t'''_{jk})/2}
(\frac{d-2}{2})_{(t''_{ik}+t''_{jk}+t'''_{ik}+t'''_{jk})/2}}
{(\frac{d}{2}+t'''_{ij}+t'''_{ik})_{s'''_i}(\frac{d}{2}+t'''_{ij}+t'''_{jk})
_{s'''_j}
(\frac{d}{2}+t'''_{ik}+t'''_{jk})_{s'''_k}}
\nonumber\\&& 
\times\frac{(\frac{d-2}{2})_{(t'_{ij}+t'_{ik}+t'''_{ij}+t'''_{ik})/2}
(\frac{d-2}{2})_{(t^v_{ij}+t^v_{jk}+t'''_{ij}+t'''_{jk})/2}
(\frac{d-2}{2})_{(t^v_{ik}+t^v_{jk}+t'''_{ik}+t'''_{jk})/2}}
{\prod 
[(\frac{d-2}{2})_{t'''_{ij}+t'''_{ik}}]^2(\frac{d-2}{2}+t'''_{ik}+t'''_{jk})
_{t'''_{ij}}}\label{eq:contr:phi}.
\end{eqnarray}
We notice that the symmetry  $l \leftrightarrow l'$ is manifest, while this
is not the case for $i \leftrightarrow j \leftrightarrow k$: we can explain this noticing that
 the summation $\sum_l$ already breaks this symmetry.
This expression can be written as $\phi=M^T D M$, with $M$ 
triangular and $D$ 
diagonal. This allows us to calculate the determinant of $\phi$:
\begin{eqnarray} 
\lefteqn{\det(\phi)=}\nonumber \\&&
\prod_l \frac{\prod s_i! t_{ij}! (d/2-t_{ij}-1)_{t_{ik}+t_{jk}}}
{(d-2)_{\Sigma t_{ij}}(d/2)_{\Sigma t_{ij}}\prod N_i! 
(d/2+t_{ik}+t_{jk})_{s_i}
(d/2+t_{ik}+t_{jk})_{t_{ij}}}.\label{eq:contr:det}
\end{eqnarray}
We have tested the correctness of expressions \Ref{eq:contr:phi} and \Ref{eq:contr:det} in a large
number of cases  by comparing with the inverse of the $\chi$-matrix generated by recurrence relations
(the number of tensor indices had to be small enough to allow the explicit
inversion done by {\tt Mathematica} without causing a hang-up).

\section{Summary and conclusions}
We have two mayor results to report. Firstly there are the general recurrence relations
for the generation of the contraction matrix which are as far as we know new in the literature.
 In itself this expression is quite useful:
for small and most common matrices explicit inversion is easy, for 
larger matrices and in all practical cases
we are solely interested in a $\varepsilon$-expansion. Since every element of 
the contraction matrix is polynomial in $d$, we can easily perform the inversion of $\chi$
perturbatively in $\varepsilon$ \cite{chet}.

Secondly in the 3-loop case we succeed in constructing a symmetrical explicit expression for 
the decomposition matrix. Apart from the esthetical satisfaction of finding a non-recurrent
solution it is also indispensable for the 
large matrix cases. 
On the other hand even for the every-day cases it is a fast and direct way to generate
the decomposition matrix, either in the case of analytic expression as a function of $d$ or
as an $\varepsilon$-expansion.

Generalisation towards more than three loops seems to be non-trivial: 
  we do not get a unique configuration in the four-loop case using the 
recurrence relations of section \ref{recur}  (e.g. 
every loop has 2 tensor indices: then there are several cases with $s_i=0$). So
we need more than a simple extension of the tric used in \cite{Davy4}.
Although we are a bit closer towards the solution of the $L$-loop problem,
a general solution for this intriguingly simply-looking problem is still lacking.

\end{document}